\documentclass{aa}
\usepackage{graphicx,txfonts,natbib,color}
\bibpunct{(}{)}{;}{a}{}{,}

\begin{document}
\title{Circumstellar disks can erase the effects of 
stellar fly--bys on planetary 
systems.}
\author{F. Marzari\inst{1}, G. Picogna \inst{1}}
\institute{
  Dipartimento di Fisica, University of Padova, Via Marzolo 8,
  35131 Padova, Italy
}
\titlerunning{Circumstellar disks and stellar fly--bys.}
\authorrunning{F. Marzari et G. Picogna}
\abstract 
{Most stars form in embedded clusters. Stellar flybys may affect 
the orbital architecture of the systems by exciting the eccentricity
and causing dynamical instability. 
}
{Since, incidentally, 
the timescale over which a cluster loses it
gaseous component and begins to disperse is comparable to
the circumstellar disk lifetime, we expect that 
closer, and more perturbing,  stellar flybys occur when the planets 
are still embedded in 
their birth disk. We investigate the effects of the 
disk on the dynamics of planets after the stellar encounter to test 
whether it can damp the eccentricity and return the 
planetary system to a non-excited state. 
}
{ We use the hydrodynamical code FARGO to study the 
disk+planet(s) system during and after the stellar encounter
in the context of evolved disk models whose superficial 
density is  
10 times 
lower than that of the Minimum Mass Solar Nebula.
}
{The numerical simulations show that the planet eccentricity, 
excited during a close stellar flyby, is damped on a short 
timescale ($\sim 10$ Kyr) in spite of the disk low initial density 
and subsequent tidal truncation. This damping is effective also 
for a system of 3 giant planets and  the effects of the 
dynamical instability induced by the passing star are quickly 
absorbed.
}
{ If the circumstellar disk is still present around the star 
during a stellar flyby, a planet (or a planetary system) is returned 
to a non-excited state on a short timescale. This does not mean 
that stellar encounters do not affect the evolution of planets, 
but they do it in a subtle way with a short period of agitated 
dynamical evolution. At the end of it, the system resumes a 
quiet evolution and the planetary orbits are circularized by 
the interaction with the disk. 
}
\keywords{Planets and satellites: dynamical evolution and stability ---
Planet--Disk interactions --- 
 --- Methods: numerical }
\maketitle
\section{Introduction} 
\label{intro}

%

It is widely believed that  most
stars form in embedded clusters on a timescale of the 
order of about 1 Myr \citep{hille97, palla00}. 
About 80\% of stars within 1 Kiloparsec of the Sun 
are in effect found in clusters with a population 
exceeding 100 members \citep{porro}.
During the early
stages of cluster evolution, stellar encounters are believed to
significantly 
affect the formation and subsequent dynamical evolution of 
planetary systems around stars belonging to the structure.  
According to \cite{malm1,malm2,malm3,zaktre}, scattering interactions with 
other stars in their birth cluster may excite the eccentricity of 
planets populating the outer parts of the system. 
This dynamical mechanism might contribute to explain 
why eccentric orbits occur 
relatively commonly in extrasolar planetary systems.
In the case of multiple-planet systems, the eccentricity 
perturbations due to a stellar flyby can leave the 
planetary system in an unstable state. On 
timescales ranging from a few millions to billions of years 
an unstable planetary system  
may undergo a phase of 
planet--planet scattering leading to the ejection of 
one or more planets from the system \citep{wema, rf96}. 

Stellar flybys are more frequent and statistically close in the 
early stages of a cluster evolution when the 
structure is more compact.
When, after a few times the crossing time \citep{allen07},   
a cluster loses its gaseous component, a substantial amount of
unbinding occurs \citep{adams00} and it disperses.  
The gas removal occurs within about 3--5 Myr from the formation 
of the cluster and, according to \cite{allen07,
fall09,chan10,duk12}, after about 10 Myr, 90\% of stars born in clusters
have dispersed into the field.
We then expect that 
most of the close stellar encounters causing significant changes 
on the orbital elements of the planets occur when the planets
are still embedded in circumstellar disks.   
Even if 
gas disk lifetimes are not well constrained observationally,
they are assumed to be shorter than 10 Myr \citep{pasc06}, possibly 
in the range between 3--6 Myr as suggested by \cite{hll06}. It is
then reasonable to assume that, statistically, most of 
the close stellar passages occur when the planets are still 
embedded in disks. The effects of stellar encounters must be 
modeled considering  
the system planet+disk rather than isolated planets. 

Indeed, most numerical modeling on the effects of stellar encounters 
on planetary systems in clusters 
are based on pure N--body simulations 
where the effects of circumstellar disks are neglected
\citep{malm1,malm2,malm3,zaktre}. 
However, as suggested by the almost coincidence between the 
cluster lifetime and that of circumstellar disks, 
this assumption may not be fully justified. Close
stellar flybys are frequent in the initial stages  
of cluster evolution when the disks are still interacting with 
the embedded planets. 
We concentrate in this paper on the effects of close 
stellar encounters on planets still embedded in their
birth disk. \cite{frane09}  have shown that 
stellar flybys significantly 
altering the orbital parameters of planets are also expected to 
affect 
the disk structure. 
Via hydrodynamical simulations they find 
that stellar encounters with distances
less than 3 times the disk radius $r_d$  can cause a significant shrinkcage 
of the disk. They concentrated on the effects of the modification 
of the disk structure on planet growth and migration suggesting that giant
planets in systems involved in stellar encounters during their early 
evolution should have higher masses and larger semimajor axes. 
We focus instead on the damping effects of the disk on the 
eccentricity of a planet after that a stellar encounter significantly
excited it. For this reason we resort to 
hydrodynamical modeling to compute at the same time the disk 
and planet evolution during and after the stellar flyby. 
Our goal is to test whether the tidal interaction of the planet(s) 
with the disk, even truncated 
after the encounter, is able to damp the planet eccentricity 
excited after the flyby. The damping might also prevent the 
onset of gravitational instability in
multi--planet systems at later times. We consider  
evolved disk 
with a density significantly lower respect to that of the 
Minimum Mass Solar Nebula since planet growth is supposed to 
have already occurred. 

The paper is organized as follows. In Sect. 2 we describe the 
numerical algorithm used to model the disk and planet evolution 
during and after an hyperbolic close encounter with another star. 
Sect. 3 is devoted to single planet evolution after a stellar flyby 
while in Sect. 4 we explore the 
evolution of 3 planet systems. 
In Sect. 5 we discuss the 
results of our numerical simulations and their implications for 
the statistical distribution of the planet orbital elements. 

\section{The numerical model} 
\label{model}

We have used the
numerical code FARGO \citep{fargo2} 
to model the time evolution of a planet embedded in 
a two-dimensional
circumstellar disk surrounding a solar type star ($ M = 1 M_{\odot}$). 
The reliability of FARGO, without additional 
special numerical resolution requirements, has been
recently confirmed by \cite{kleFA12}.
The code solves the hydrodynamical equations on a polar
grid, and it uses an upwind transport scheme along with a harmonic,
second-order slope limiter \citep{vl77}.
We focus on evolved disks since 
we assume that planets have already formed or are in their final growth
stage. At this stage, along with viscous diffusion, the disk is
also dispersed by photo--evaporation produced by photons emitted by 
the central star and, possibly, by nearby stars. 
For this reason we adopt a low initial disk's surface density
which allow us to neglect the effects of disk self-gravity. For the 
same reason we can
work in the assumption that the disk is locally isothermal so that
the temperature depends only
on r and  H/r is a constant. The initial density profile is 
$\Sigma = \Sigma_0 r^{-1/2}$ where $\Sigma_0$ is the 
2D density at 1 AU from the star. $\Sigma_0$  is set to $100 g/cm^2$, 
a value significantly lower than that of the Minimum Mass Solar Nebula
which is at least ten times higher,
while for the aspect ratio H/r we choose 0.05. A
constant shear kinematic viscosity, $\nu = 10^{-5}$ in code units
(mass unit is $1 M_{\odot}$, $G$ is equal to 1 while the length unit 
is set to 1 AU),
is adopted in all simulations. The disk ranges initially from 
0.5 to 30 AU and the density is smoothly reduced to a floor value of 
$1 \times 10^{-9}$ (in code units) beyond 30 AU. The computational domain 
extends to 50 AU and is discretized in $864 \times 240$ grid zones, in r,
and $\theta$. 
An outflow boundary condition is adopted
at both the grid's inner and outer edges.
No mass can flow back into the system
once it escaped.
The secondary star ($ M = 1 M_{\odot}$) is started on a hyperbolic orbit 
having a minimum impact paramater $q$ fixed at the beginning
of the simulation, larger than the outer
border of the grid.  It is initially located at a distance of 800 
AU from the star with the planetary system and the relative 
velocity at infinity is set to 1 km/s, a typical value in clusters. 

One or more planets are considered and their orbits are affected
by the disk, the mutual gravitational attraction if more
than one planet are present, 
the gravitational force of the central star and that
of the passing by star.  
To properly model the strong gravitational interactions between 
the passing star and the planets, 
the
numerical integrator computing the planet orbits (a 5th order Runge
Kutta in FARGO) has been updated
with a variable stepsize.

\begin{figure}[hpt]
\resizebox{90mm}{!}{\includegraphics[angle=0]{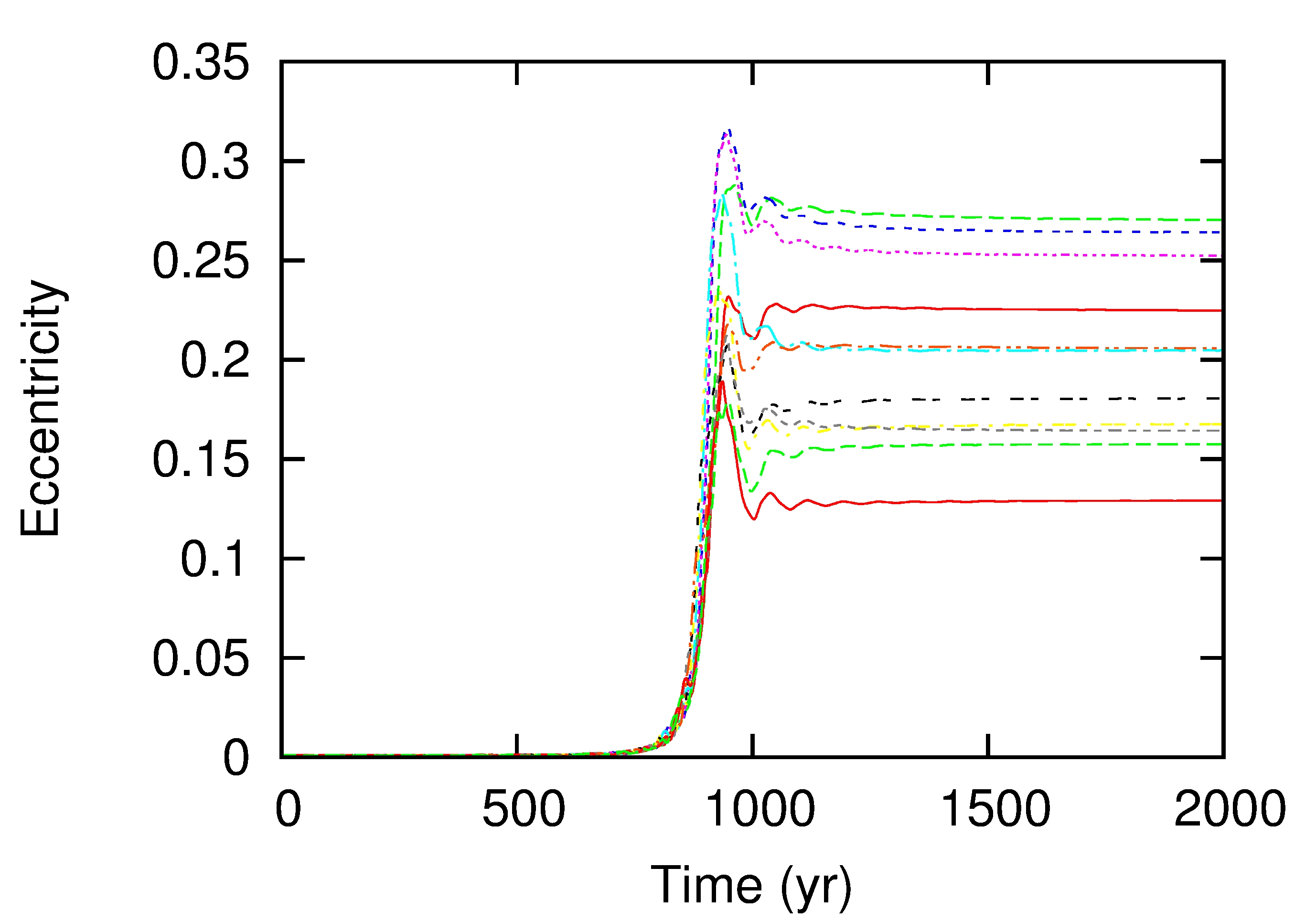}}
   \caption{\label{fph1} Evolution of the eccentricity 
of a Jupiter size planet during and after 
a close
stellar encounter in a pure 3--body problem (no disk). 
The initial mean anomaly $M$ of the planet is sampled 
with a step of $30^o$ starting from $0^o$. The 
eccentricity step $\Delta e$ depends on the initial 
value of $M$ which changes the geometrical configuration 
of the 3 bodies at the stellar encounter. 
     }
\end{figure}

The orbital changes of the planet orbit caused by 
the stellar flyby significantly depend on the 
approach configuration. In Fig.~\ref{fph1} we show the outcome 
of pure 3--body numerical integrations (the disk is neglected) 
showing the eccentricity evolution of the planet for 
different initial values of its mean anomaly $M$ sampled
from $0^o$ to $360^o$ with a constant interval of 
$30^o$ 
(the second star trajectory is kept fixed).  
Since the initial orbit of the planet is circular,
this is the only relevant angle in determing the relative position of 
stars and planet when the flyby occurs. The jump in 
eccentricity $\Delta e$ due to the stellar flyby perturbations 
significantly depends on $M$ and ranges approximately 
from about 0.13 to 0.27. Since our goal is to show that 
the disk is able to damp the eccentricity pumped up 
during stellar encounters, in our full model 
stars-planet-disk we always look for initial conditions 
leading to the largest $\Delta e$. However, 
the outcome shown in Fig.~\ref{fph1} give us only an indication 
of the eccentricity variation but cannot guide us 
in looking for the initial value of $M$ to set in the 
input file of FARGO to get the maximum $\Delta e$. This because 
the evolution of the planet embedded in the disk 
is different from a pure 3--body problem. The interaction 
with the disk causes migration in the time interval from  the 
beginning of the simulation and the occurrence of the stellar
encounter and this leads to a different geometrical configuration 
at the encounter. For this reason, in each model we perform 4 
test runs where we sample the initial value of $M$ of the planet
and we continue the model with the largest $\Delta e$,
 the most perturbing configuration. We cannot perform 
an accurate sampling like that shown in Fig.~\ref{fph1} since 
running the full model planet--stars--disk is much more 
time consuming. 

The FARGO code is two-dimensional (2D), so our modeling covers 
those cases where the trajectory of the incoming star is 
not very inclined respect to the planet orbit. A 
full three-dimensional (3D) approach
would be able to cover a wider range of situations where the 
inclination of the passing star is large. In this case we expect
not only a significant warping of the disk but also a step 
in the inclination of the planet in addition to that in eccentricity. 
However, once the warping of the disk is damped after the 
stellar encounter, the inclination of the disk relative 
to the new disk plane might be quickly damped as 
shown in \cite{marne09}, \cite{klebi11} and \cite{cres07}.
As a consequence, we expect that also
in an inclined configuration the final outcome would be a
damping of both the eccentricity and inclination forced by
the stellar flyby and then an almost complete cancellation of 
the flyby effects on the planetary system in terms 
of eccentricity and inclination excitation. 
We plan to perform in the future 3--D
simulations to explore the detailed evolution of the system
when a significant inclination is assumed for the star
trajectory
but it would be a more
difficult task due to still very time-consuming performance of
3D codes.

It is interesting that \cite{cres07} in their paper 
compared the damping 
rate of eccentric Neptune size planet in 
2D and 3D models 
in denser
disks and then less evolved
compared to those we study in this paper. They conclude that 
for large initial eccentricities of the planet orbit ($e > 0.1$) 
there is a discrepancy 
in the eccentricity damping rate of about 20\% possibly due to the 
potential softening applied in 2D simulations. Their scenario is 
different from ours since they have an almost stationary 
disk perturbed only 
by the planet while our disk is strongly affected by the 
stellar flyby. 
However, this is an indication that 3D simulation would confirm 
our results concerning the eccentricity damping of the planetary
system after the stellar encounter, even if at a different rate. 
This is not a potential problem since the timescale of the 
eccentricity damping is short compared to the lifetime of the 
leftover disk in our scenario.

\section{Case A: a single planet orbiting at 18 AU from the star} 
\label{case1}

We first consider a planet with a mass equal to $M_p = 30  M_{\oplus}$,
a super--Earth or the core of a giant planet, 
initially on a 
circular orbit ($e = 0$)
around the star with semimajor axis $a = 18 AU$.  
Such a large semimajor axis is adopted to cover the 
extreme cases where the planet is located close to
the outer border of the disk after the stellar 
flyby and might be less affected by 
the disk force. When the planet is well 
embedded in the disk, it is expected that the damping 
is more fast and efficient. 
The minimum approach distance during the stellar encounter 
$q$ is set to 70 AU. This is a close encounter configuration 
that has strong effects on both the disk and planet orbit. 
In fact, when the passing star approaches 
the system, a significant 
amount of mass is stripped away from the disk. At the same time, the orbital 
elements of the planet are changed on a short timescale. A sudden 
jump occurs in both eccentricity and semimajor axis, as predicted 
by  \cite{malm1,malm2,malm3,zaktre}.

\begin{figure*}[hpt]
\resizebox{100mm}{!}{\includegraphics[angle=-90]{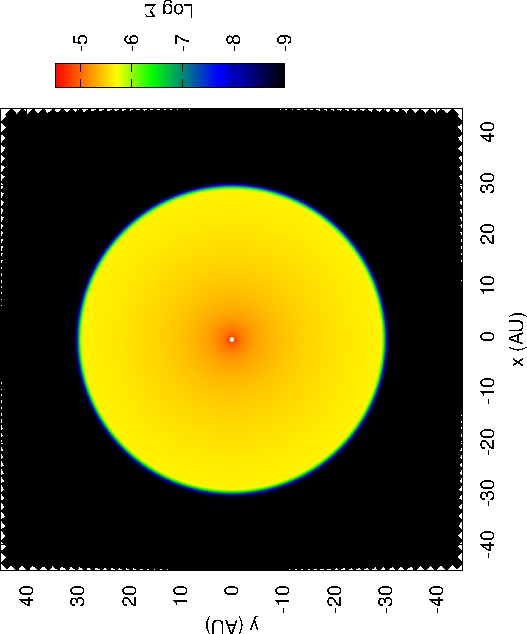}}
\resizebox{100mm}{!}{\includegraphics[angle=-90]{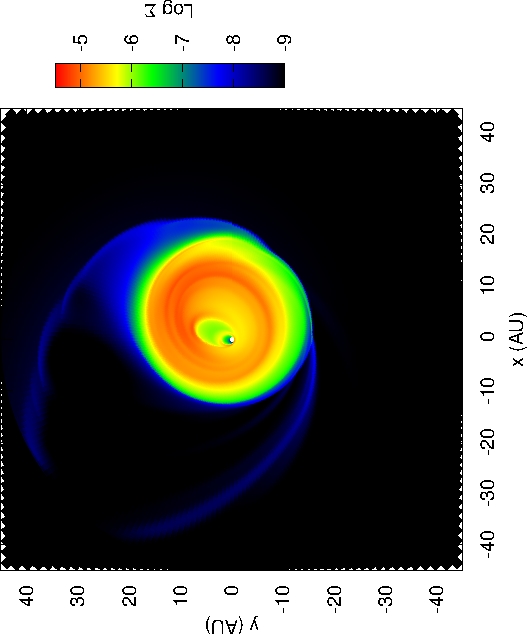}}
\resizebox{100mm}{!}{\includegraphics[angle=-90]{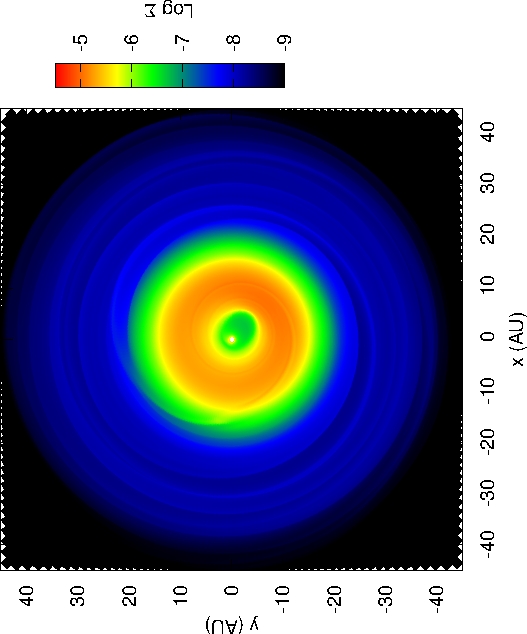}}
\caption{\label{fd1} In panels 1,2,3 we show the contours of the gas 
     density (in normalized units) when the passing star is at 800 AU 
     before the pericenter passage, 150 AU and 10000 AU after the pericenter, 
     respectively.
     }
\end{figure*}

In Fig.~\ref{fd1} we show the evolution of the disk during the 
stellar flyby. More than half of the initial mass  
$M_{d0} = 0.008 M_{\odot}$ is lost and the disk is left with 
$M_d = 0.003 M_{\odot}$. Just after the encounter the disk is
shrinked to about 12 AU and it relaxes to 14 AU with time 
(assuming that its border is marked by a density of 
$10^{-5.5}$ which is the lowest density of the disk 
at the truncation radius and  
before the stellar encounter). The elliptical 
internal low density region around the central 
star has been observed in a number of previous numerical
studies of isothermal disks in binaries \citep{kle08,marba1,marba2}.
It is related to the formation of strong spiral density
waves that propagate all over the disk and cause a flow of mass 
through the inner border of the disk down to the star surface. 
It disappears at later times due to the continuing viscous evolution
of the disk. 

\begin{figure}[hpt]
\resizebox{90mm}{!}{\includegraphics[angle=0]{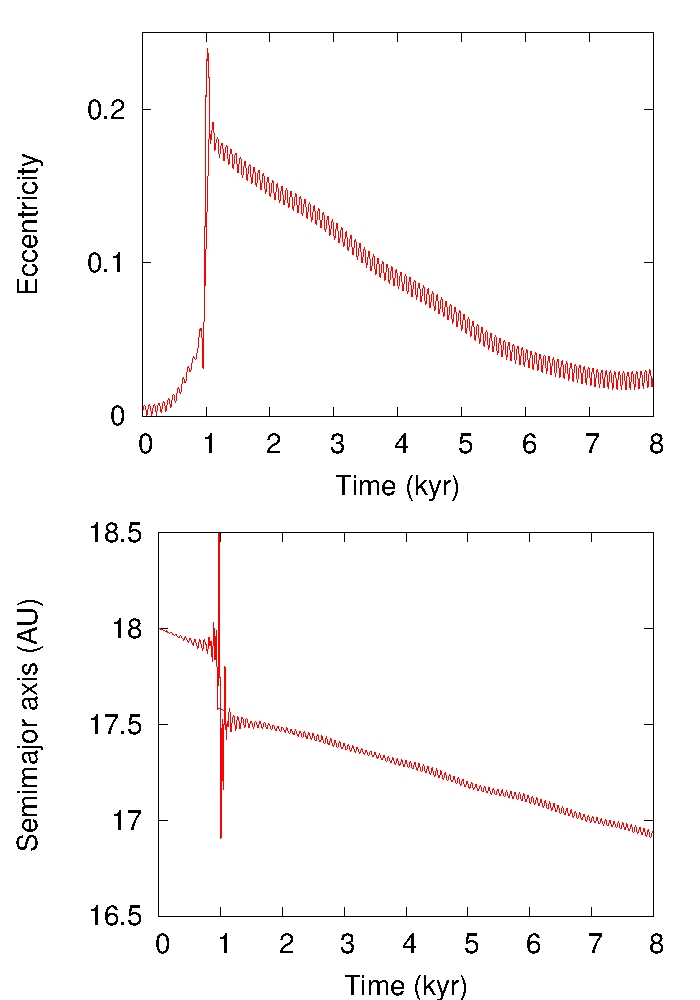}}
   \caption{\label{fp1} Evolution of the eccentricity and 
semimajor axis of a light planet ($M_p = 30  M_{\oplus}$) 
in a circumstellar disk during and after  a close
stellar encounter. The eccentricity is damped on a short timescale
and normal migration is resumed. 
     }
\end{figure}

In spite of the strong mass
depletion, the disk is still able to interact with the planet
and circularize its orbit. This behaviour is illustrated in
Fig.~\ref{fp1}.  The planet eccentricity is excited to 
about 0.2 and the semimajor axis jumps down to 17.5 AU
during the interaction with the passing by star.
However, the subsequent interactions between the disk and 
the planet damps the eccentricity on a short timescale 
(less than $10^4$ yr) and, at the same time, the planet
resumes its type I migration inwards. It is clear from this 
example that planets on inner orbits, if excited by 
stellar encounters, would return to circular orbits 
on an even shorter timescale since the gas density increases 
closer to the star. 

A second simulation was performed with a Jupiter size planet
$M_p = 1  M_{J}$ and the same configuration for the 
disk and the passing star. To make the model more precise 
we should have run the code without the passing star and 
with the planet on a fixed orbit to give it the time to 
carve a gap in the disk at its location prior to the 
stellar flyby. However, as also 
shown by \cite{frane09}, a close stellar encounter 
strongly perturbs the disk destroying any previous structure 
present in it. A pre-existing planetary gap would be fully 
erased by the tidal perturbations of the passing star. 

\begin{figure}[hpt]
\resizebox{90mm}{!}{\includegraphics[angle=0]{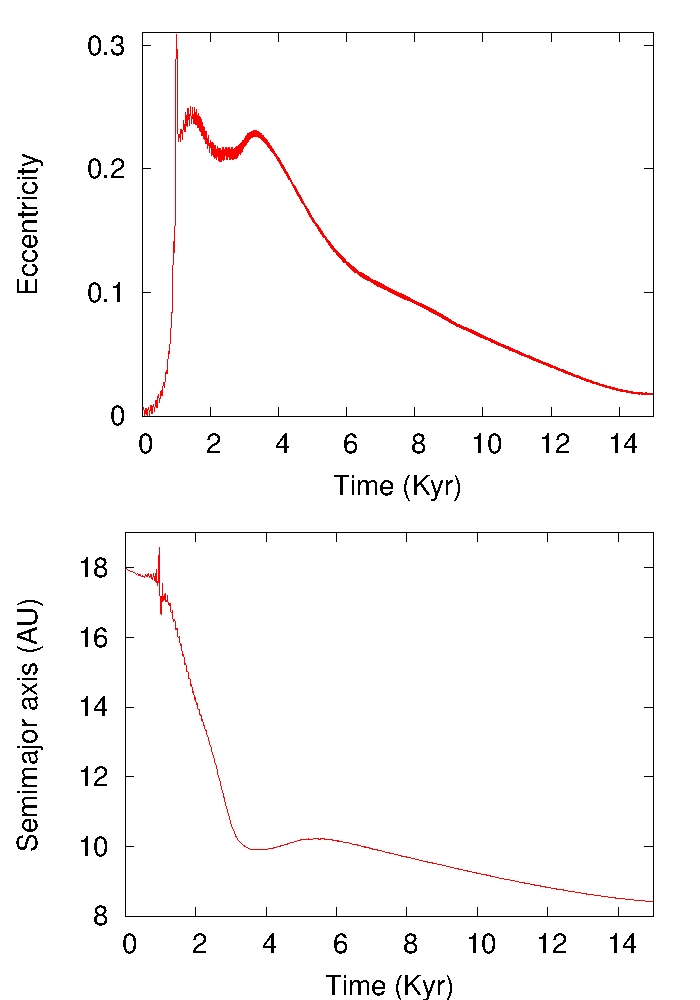}}
   \caption{\label{fp2} Same as in Fig.~\ref{fp1} but for a 
more massive planet ($M_p = 1  M_{j}$). Even in this case 
the eccentricity is damped and type II migration resumed. 
     }
\end{figure}

As in the case of the light planet, the eccentricity of the 
Jupiter size planet, after the initial step due to the passing
star perturbations, is damped on a short timescale as shown 
in Fig.~\ref{fp2}. Just after the stellar encounter the planet 
is left on an eccentric orbit and without a gap. It
has a fast inward migration which halts when its eccentricity 
is damped to almost 0. At that time, it has carved a new gap around its 
orbit and it resumes its regular type II migration. Even in this 
case the stellar encounter does not leave the system with a 
planet on an eccentric orbit, but it eventually causes a period 
of rapid inward migration after destroying the gap that the planet 
developed prior the encounter. But, after about $10^4$ yr, the 
system has absorbed most, if not all,  the perturbing effects of the 
stellar encounter. The disk may still bear some eccentricity but 
the viscous evolution will bring it back to a circular 
state. When this happens, the system will have erased all records 
of the close stellar passage. 

\section{Case B: 3 planets orbiting the star} 
\label{case1}

Stellar encounters are important for multi--planet systems since 
they could trigger a phase of dynamical instability followed by the
ejection of one or more planets and significant changes in the 
orbital architecture. If  
the eccentricity of the outer planet is 
excited by a passing star, Hill's stability might be destroyed 
leading to a 
phase of planet--planet scattering at later times
\citep{malm1,malm2,malm3,zaktre}.  A large eccentricity is in 
effect the first step to the development of a chaotic behaviour. 

We explore here the case of a system made
of 3 Jupiter size planets embedded in an evolved disk and  
migrating towards the star. We preferred a 3--planet system 
rather than a 2--planet one for its higher dynamical 
complexity and stronger tendency to 
develop crossing orbits \citep{chat}. \cite{mar10} 
already studied a system of 3 giant planets embedded in 
a circumstellar disk and they showed that the migration 
leads the planets into different evolutionary paths, either 
mutual resonance trapping or planet--planet 
scattering. The choice between the two dynamical paths depends 
on the masses of the planets and on the disk physical properties. 
In this paper we investigate if the stellar flyby is 
always leading to planet--planet scattering with final 
eccentric orbits that can be observed or if, again, the 
influence of the disk is able to damp the eccentricity 
of the planets and erase any memory of the chaotic 
phase.

\begin{figure}[hpt]
\resizebox{90mm}{!}{\includegraphics[angle=0]{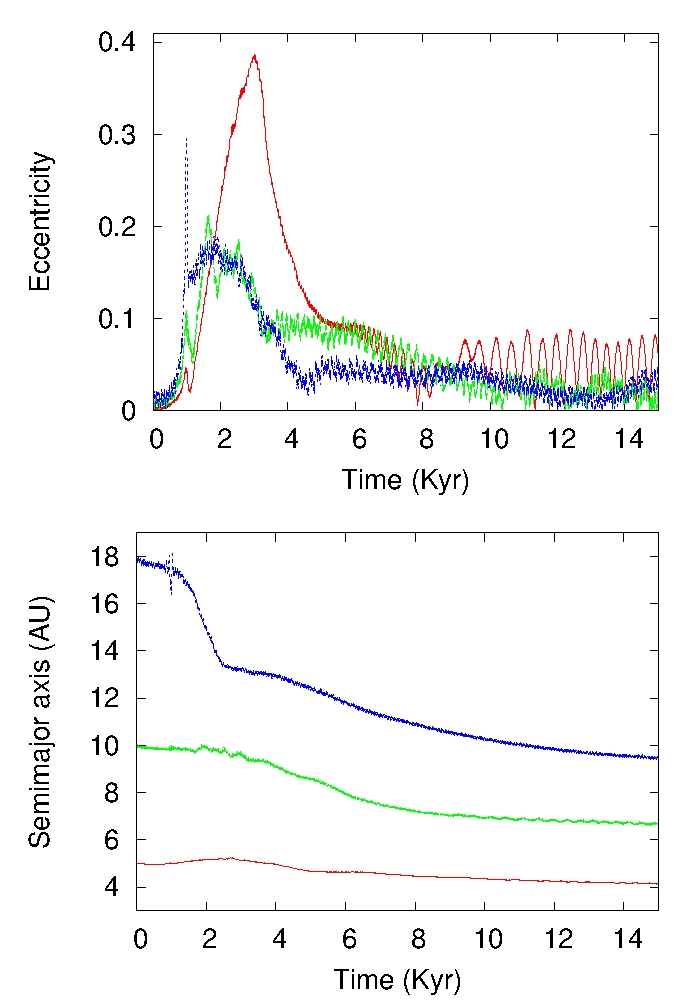}}
   \caption{\label{fp3-1} Evolution of a 3--planet system 
($M_{p1} = M_{p2} = M_{p3} = 1  M_{j}$) during and after
the stellar encounter. Even in this case the eccentricity
is damped in less than 10 Kyr. 
     }
\end{figure}

In Fig.~\ref{fp3-1} we illustrate the evolution of a 3--planet
system with the planets initially on circular orbits with 
semimajor axis 5,10,18 AU, respectively. The eccentricity of 
all planets are excited during the stellar encounter. The outer
planet shows a sudden step when the passing star reaches the 
pericenter and the other two planets are strongly perturbed,
at subsequent times, 
both by the eccentricity of the outer planet and by the strong 
eccentricity developed by the disk. The inner planet is the most
affected by the turmoil of the stellar encounter and its eccentricity
is slowly pumped up to almost 0.4.
However, the disk is slowly circularized and the subsequent 
disk--planet interactions damp the eccentricities of all the planets.
After about $10^4$ yr, the system has completely erased 
the effects of the stellar encounter and all the three planets migrate
towards the star at a slow rate trapped in a mutual 
2:1 resonant configuration. Their further evolution 
would be no different from that observed in a single star 
which did not undergo a stellar encounter even if a longer 
integration would be necessary to figure out if the resonant
capture is stable or not. However, this is out of the scope
of this paper focused on showing that the disk is able to 
erase the effects of stellar encounters.

\begin{figure}[hpt]
\resizebox{90mm}{!}{\includegraphics[angle=0]{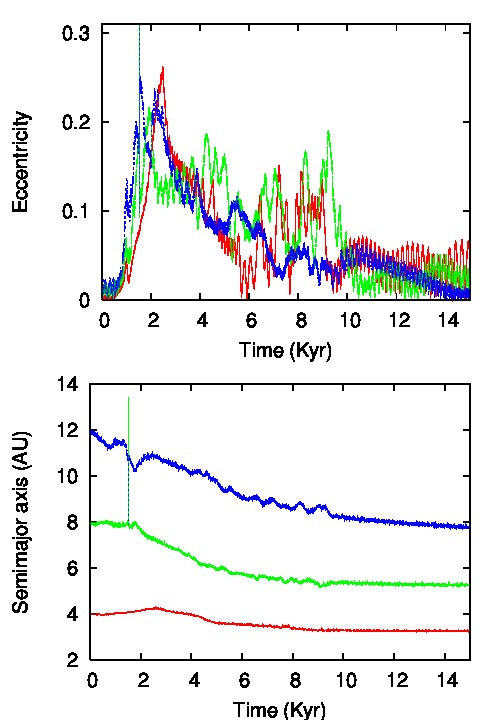}}
   \caption{\label{fp3-2} Same as in Fig.~\ref{fp3-1} but for
a system initially mode compact 
($a_{1} 4 AU$, $a_{2} = 8 AU$ and $ a_{3} = 12 AU $). The eccentricity is 
damped and only the forced component due to mutual gravitational
interactions between the planets is left. 
     }
\end{figure}

\begin{figure}[hpt]
\hskip -2 truecm
\resizebox{130mm}{!}{\includegraphics[angle=-90]{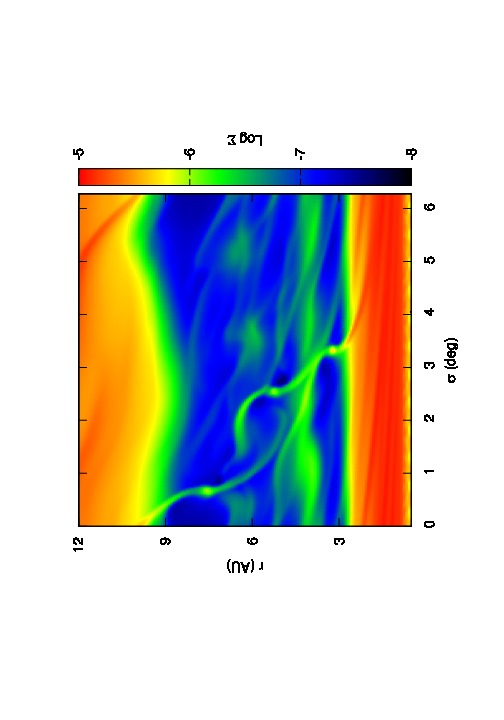}}
   \caption{\label{f3hy-1} Disk density distribution in 
cylindrical coordinates after the stellar flyby. 
The three planets are migrating insede and have carved a partial 
gap. The two inner planets are in a 2:1 mean motion resonance. 
     }
\end{figure}

A second more compact system has been modeled where the planets 
are closer to each other. The inner one has a semimajor axis of 
4 AU, the middle one of 8 AU and the outer of 12 AU. The stellar 
hyperbolic trajectory is kept the same. Even in this case we
observe a consistent eccentricity excitation for all three
planets. 
In Fig.~\ref{fp3-2} we show the orbital element 
evolution of the system before and after the close stellar passage. 
A close encounter occurs between the two outer planets, 
but even in this case on the long term the disk damping takes over and 
the eccentricity is slowly reduced after about 10 Kyr from the 
stellar encounter.
The system returns to a quiet evolution with the planetary 
eccentricities reduced to pre--encounter values 
with an additional forced component due to the mutual 
secular perturbations. The planets keep migrating 
inside within a common gap as shown in 
Fig.~\ref{f3hy-1} . The two inner planets are trapped in a
2:1 resonance while the outermost planet is drifting at almost the 
same speed as the resonant pair.  This does not 
mean that the stellar encounter did not affect the evolution 
of the system. In effect, the semimajor axes possibly evolved 
faster during to the period of high eccentricity. However, at the 
end of the excited period, the eccentricities are damped 
and, simply by looking at the values of the semimajor
axes, it is not possible to guess the occurrence of the 
stellar encounter. This same behaviour was observed in other simulations
where the initial orbital elements had different initial 
orbital angles. 

As shown in \cite{mar10}, 
a large variety of outcomes are possible after the chaotic scattering 
phase ranging from orbital exchange, planet merging and scattering of a 
planet in a hyperbolic orbit. Of course in 2D models like 
those shown in \cite{mar10} and presented here the 
probability of planet merging is 
significantly increased respect to a more realistic 3D model 
since the 2D cross section for impact is significantly larger 
respect to 3D one. However, it is a possibility and it 
needs to be taken into account. 
In the case of 3 planet systems perturbed by a stellar flyby we 
expect that the statistical distribution of all possible outcomes 
to be different from that in the unperturbed case. In particular, 
the occurrence of planet--planet scattering like that shown 
in Fig.~\ref{fp3-2} may occur more frequently compared to the 
unperturbed case due to the large perturbations in the 
planet orbits induced by the stellar encounter. 
However, a 
deep exploration of the parameter space is required and this
appears to be a really complicated task due to the large number of free
parameters like the masses of the planets, their initial orbits
and the disk parameters. Undertaking this task would outline 
the real influence of stellar flybys on the planet architecture 
of systems embedded in clusters, not in terms of 
eccentricity excitation but in the 
semimajor axis and mass distribution of the planets. 
Unfortunately, the required 
amount of CPU time even for a 2D statistical 
exploration appears at the moment 
out of reach.  

\section{Resolution issue}
\label{reso}

\begin{figure}[hpt]
\resizebox{90mm}{!}{\includegraphics[angle=-90]{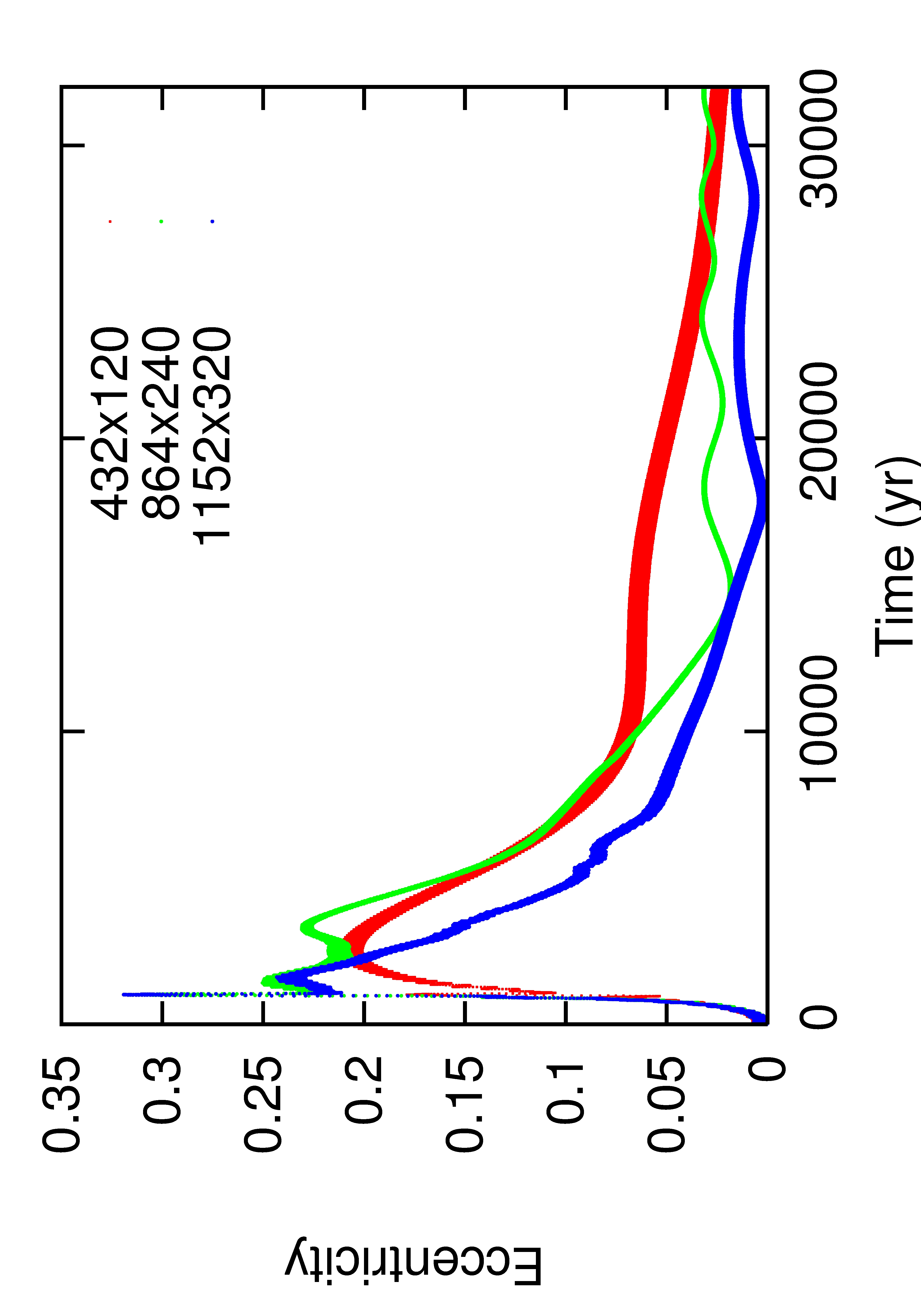}}
   \caption{\label{fr1} 
    Eccentricity evolution of a single massive planet for 3 different 
    grid resolutions: 432x120, 864x240 and 1152x320, respectively. 
    The parameters are the same used in the model shown in 
    Fig.~\ref{fp2}. 
    }
\end{figure}

To validate our results and demonstrate that the damping of 
eccentricity after a stellar flyby is a robust result 
and it does not depend on the 
grid resolution used in 
FARGO, we performed two additional simulations for the case
of a single massive planet shown in Fig.~\ref{fp2} which represents
our standard model. The first is a lower resolution run
with a grid size of 432x120 while the second is 
a high resolution run with a 1152x320 grid. The outcome of these
two additional models are shown in Fig.~\ref{fr1} where they 
are compared to the standard resolution model. The three runs 
confirm that the eccentricity damping occurs independently of 
the resolution used in the code. However, there are differences 
in the eccentricity evolution which are mostly related to the 
configuration of the planet at the stellar encounter. As shown in 
Fig.~\ref{fph1}, the eccentricity jump depends on the 
position of the planet in its orbit, i.e. on its mean anomaly 
$M$,  when the stellar passes close 
to the planet. If a different resolution is used, prior to the 
stellar encounter the evolution of the semimajor axis due to the interaction 
with the disk will be slightly different and the value of
$M$ at the stellar encounter will differ in the three runs. 
This leads to an avalanche 
effect which affects the peak eccentricity at the stellar 
flyby, the subsequent planet migration and the damping rate. 
However, Fig.~\ref{fr1} confirms that the erasing of the 
stellar flyby effects is a robust effect and the resolution is 
not altering the morphology of the long term evolution of the
system which tends towards the circularization of the planet
orbit. 

\section{Discussion and conclusions}
\label{end}

We have shown that circumstellar disks are able to absorbe the 
effects of close stellar encounters on planetary systems 
orbiting stars in clusters. The orbital eccentricity of 
planets is excited during the stellar passage, but the 
disk damps it back to low values on a short timescale 
of the order of 10 Kyr. This damping is very efficient 
since in our simulations we consider evolved disks with 
a density at least 10 times lower than the MMSN. This 
assumption is dictated by the fact that planets take time 
to form and in the meantime the disk is slowly dissipating by 
viscous evolution and photo--evaporation. 

The relevance of disk eccentricity damping may be significant
since stellar flybys are expected to be more frequent 
and closer in the early stages of a stellar cluster lifetime. 
Incidentally, the timescale over which a cluster loses its
gaseous component and begins to disperse is comparable to 
the circumstellar disk lifetime. Statistically, a large fraction of 
close stellar encounters are expected to occur while the circumstellar disk 
is still present and able to damp the eccentricity induced 
by the stellar flyby.  

Our results do not imply that stellar flybys do not affect 
the evolution of planetary systems in clusters. However, they may do 
it in a more subtle way if the circumstellar disk is 
still present. The eccentricity is quickly damped, but 
the evolution of the system, in particular planet migration, is 
faster when the eccentricity is excited and the disk 
may also temporarily enhance the eccentricity excitation, 
as shown in the case of 3 planet systems. Even close encounters
can occur, but after the period of dynamical excitation, the 
disk damps the eccentricity and the system returns to a 
quiet state and it resumes a normal migration speed. 
By inspecting the dynamical properties and 
architecture of planetary
systems around stars that were members of clusters, it would be 
be difficult 'a posteriori' to distinguish 
between systems whose evolution was influenced by stellar flybys 
and those that were not. The influence of stellar flybys will be 
detectable only on a statistical base as shown by the 
modeling of \cite{frane09}. Many parameters are indeed affecting the 
behaviour of the disk+planet system during the stellar
flyby like the initial disk density profile and 
the architecture of the planetary system. A large number of 
simulations is required to statistically asses the contribution 
of stellar flybys to the evolution of planetary systems in
clusters. 

Our modeling is based on a 2D code, so the results apply 
to scenarios where the passing by star has a small 
inclination respect to that of the planet orbit. In the future 
we plan to perform full 3D simulations to test the effect 
of an inclined stellar flyby on the planetary system embedded
in the disk both on the eccentricity and inclination of 
the planet.

\section*{ACKNOWLEDGMENTS}

\bibliographystyle{aa}
\bibliography{hyper}

 \end{document}